\documentclass[12pt]{article}
\usepackage{amsfonts}
\usepackage{amsmath}
\usepackage{epsfig}

\begin{document}

\begin{flushright}
01/2015\\
\end{flushright}
\vspace{20mm}
\begin{center}
\large {\bf A microscopic Interpretation of the SM Higgs Mechanism}\\
\mbox{ }\\
\normalsize
\vspace{1.0cm}
{\bf Bodo Lampe} \\              
\vspace{0.3cm}
II. Institut f\"ur theoretische Physik der Universit\"at Hamburg \\
Luruper Chaussee 149, 22761 Hamburg, Germany \\
\vspace{3.0cm}
{\bf Abstract}\\
\end{center} 
A model is presented where the Higgs mechanism of the Standard Model is deduced from the alignment of a strongly correlated fermion system in an internal space with $A_4$ symmetry. The ground state is constructed and its energy calculated. Finally, it is claimed that the model may be derived from a field theory in 6+1 dimensions.



\newpage

\normalsize



\section{Introduction}







The Higgs sector of the Standard Model (SM) of elementary particles and the associated spontaneous symmetry breaking (SSB) show a strong similarity with the Landau-Ginzburg description of superconductivity as well as with the linear sigma model of pion physics, and it has long been speculated, that just as in those cases an underlying microscopic pairing interaction may be at work in the SM. One option put forward already in 1979 is that the Higgs particle may be composed of 'techniquarks' U and D\cite{wein1,suss1}, in a similar way in which pions are composed of up- and down-quarks u and d, and a technicolor QCD-like theory was suggested for the underlying dynamics. The main drawback of such technicolor models, in particular in their 'extended' form, is the appearance of unwanted flavor changing neutral currents (FCNC)\cite{technireview}.

The starting point of the present approach is an isospin doublet $\psi=(U,D)$ of Dirac fermions reminiscent to technicolor models, however without a technicolor quantum number and, to avoid FCNCs, without a direct interaction to quarks and leptons. Rather we shall assume that the pairing mechanism is due to exchange interactions and strong correlations of fermions, effects which in many body physics are known to be responsible for SSB in superconductors and (anti)ferromagnets. In contrast to solid state physics we do not consider these effects in physical space, but attribute them to arise from an independent dynamics which is active in the internal spaces. To be concrete, we assume the existence of a non-relativistic real internal 3-dimensional space $R^3$ with rotational SO(3)-symmetry for which the doublet $\psi=(U,D)$ serves as an (internal) Pauli spinor with an initial internal SU(2) spin symmetry. The geometrical picture is that the world is a fiber bundle over Minkowski space with fibers given by the $R^3$ spaces, and that within these fibers physical processes take place. We further assume that at high temperatures there is a symmetric state in which the internal spins are distributed randomly in the fibers, giving rise to a local SU(2) symmetry of the Lagrangian, local in the sense that on each site in each fiber the spins may be rotated independently. With respect to Lorentz symmetry both U and D can appear as lefthanded or righthanded objects, so that one may in fact consider separately a $SU(2)_L$ for the lefthanded and $SU(2)_R$ for the righthanded objects.



To recapitulate, the Standard Model SSB is triggered by the Higgs field H, a doublet under $SU(2)_L$ which via a symmetry breaking potential 
\begin{eqnarray}  
V(H)=-\mu^2 H^+ H + \lambda (H^+H)^2 
\label{hipo}
\end{eqnarray}
acquires a non-vanishing vacuum expectation value $\langle H^+ H \rangle =\frac{\mu^2}{2\lambda}$. More in detail the Higgs doublet can be parametrized as 
\begin{eqnarray}
H=\frac{1}{\sqrt{2}}
\begin{pmatrix}
i(\pi_x-i \pi_y) \\
\sigma -i\pi_z
\end{pmatrix}
\label{hig77}
\end{eqnarray}
so that 
\begin{eqnarray}  
V(H)= -\frac{1}{2}\mu^2 (\sigma^2 +\vec\pi^2)+\frac{1}{4}\lambda (\sigma^2 +\vec\pi^2)^2
\label{hipo2}
\end{eqnarray}
with minimum at 
\begin{eqnarray}  
\Lambda_F^2:= \langle \sigma^2 \rangle =\frac{\mu^2}{\lambda}
\end{eqnarray}  
which is often called the Fermi scale. 
Note that $\sigma$ is a real scalar field, while $\vec \pi=(\pi_x,\pi_y,\pi_z)$ is an axial vector field which can be interpreted as the longitudinal components of the afterwards massive W/Z bosons. In the framework of our model $\vec \pi$ can be identified with the internal chiral spin vector, and x, y, z are the coordinates of the internal 3-dimensional $R^3$ space.  

Although $\pi$-condensates could be conceivable, in particle physics it turns out that the vev is attributed to the $\sigma$ field alone, i.e.
\begin{eqnarray}
\langle H \rangle =
\frac{1}{\sqrt{2}}
\begin{pmatrix}
0 \\
\langle \sigma  \rangle  
\end{pmatrix}
=\frac{1}{\sqrt{2}}
\begin{pmatrix}
0 \\
\Lambda_F
\end{pmatrix}
\label{hvev}
\end{eqnarray}
The shifting relation $\sigma=\Lambda_F+\phi$ defines the physical Higgs particle $\phi$, whose tree level mass can easily be shown to be $m_\phi = \sqrt{2} \mu$. The values $\Lambda_F=246$ GeV and $m_\phi=124$ GeV fix the Higgs potential completely.
\begin{figure}
\begin{center}
\epsfig{file=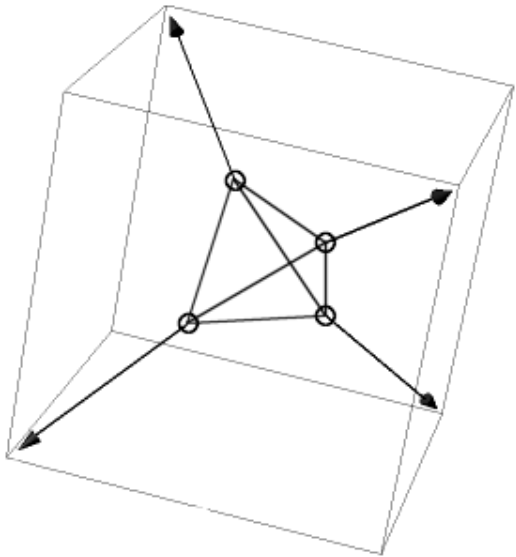,height=9.5cm}
\caption{The local ground state of the model, living in a 3-dimensional internal $R^3$ space (called the 'fiber'). Shown are the corner points (small circles) of the internal tetrahedron, which can be represented by their coordinate vectors $\vec r_i$. The origin of coordinates is taken to be the center of the tetrahedron, and is identical to the base point of the fiber in Minkowski space. 
On each corner point $i=1,2,3,4$ there is a chiral spin vector $\vec \pi_i$, pointing in the same radial direction as $\vec r_i$. (Note that the spin vectors are shown but not the coordinate vectors $\vec r_i$.) The tetrahedron itself has the tetrahedral group $S_4$ as point group symmetry. However due to the pseudovector property of the spin vectors the whole system has the Shubnikov point symmetry $A_4 + S ( S_4 - A_4)$\cite{shub}, where S is the internal time reversal operation and $A_4$ is the subgroup of $S_4$ which does not contain reflections. The Shubnikov group is chiral, the configuration with opposite chirality being given when the 4 spin vectors would point inwards instead of outwards. Before the formation of the chiral tetrahedron the internal spins U and D, which according to eq. (\ref{multi1}) are the building blocks of the spin vectors $\vec \pi_i$, can freely rotate and thus there is an internal spin SU(2) symmetry group, which however is broken to $A_4 + S ( S_4 - A_4)$ when the chiral tetrahedron is formed.}
\nonumber
\end{center}
\end{figure}

\section{Symmetry Breaking in an $A_4$ model}



In ref. \cite{lamm1} it was shown that the internal (spin and vibrational) excitation spectrum of the Shubnikov group $A_4 + S ( S_4 - A_4)$\cite{shub,bata,borov,alta} yields the correct multiplet structure of all 24 quark and lepton states of the 3 families 
\begin{eqnarray} 
A(\nu_{e})+A'(\nu_{\mu}) +A''(\nu_{\tau}) +T(d)+T(s)+T(b)+ \nonumber \\
A_s(e)+A'_s(\mu)+A''_s(\tau) + T_s(u)+T_s(c)+T_s(t) 
\label{eq833hg}
\end{eqnarray}
where $A$, $A'$, $A''$ and $T$ are singlet and triplet representations of $A_4$ and the index s denotes genuine representations of the Shubnikov group\cite{shub}. It should be stressed that the eigenmodes of the vibrating spin vectors lead to exactly these 24 states, not less and not more, and arranged in such a way that the correct mass spectrum naturally arises.

This discovery has led to the main assumption of the present paper, namely that fig. 1 should be taken as the {\it local} ground state of the model. In other words, it is assumed that in each of the 3-dimensional internal $R^3$ fibers there is a discrete tetrahedral structure and that the internal dynamics is such that spin vectors arrange themselves according to this internal tetrahedral symmetry, as depicted in fig. 1. Shown are the corner points of the internal tetrahedron and on each corner point $i=1,2,3,4$ the chiral spin vector $\vec \pi_i$, pointing in the same radial direction as the coordinate vector $\vec r_i$. (The $\vec r_i$ are not shown in the figure, and the precise mathematical definition of the chiral spin vectors $\vec \pi_i$ will be given later in eq. (\ref{multi1}).) The tetrahedron itself has the tetrahedral group $S_4$ as point group symmetry. However, due to the pseudovector property of the internal spin vectors the whole system loses its reflection symmetries and obtains instead the Shubnikov symmetry group $A_4 + S ( S_4 - A_4)$\cite{shub,bata,borov}, where S is the internal time reversal operation and $A_4$ is the subgroup of $S_4$ which does not contain reflections. Note that S itself does not belong to the Shubnikov group, and also the internal reflections do not. In other words, both internal parity Q and time reversal S are violated by the ground state fig. 1. Only their product SQ is a symmetry of the system. One can rephrase this by stating that the ground state and its symmetry are chiral with respect to the internal coordinates, the configuration with opposite chirality being given when the 4 spin vectors would point inwards instead of outwards. 

In section 1 it was argued that the internal $R^3$ spaces are nonrelativistic, at rest (no boosts allowed, because they are fixed to their base point in Minkowski space) and rotationally invariant, with an internal rotational SO(3) and a corresponding spin SU(2) under which the fundamental spinor $\psi=(U,D)$ transforms. Due to this symmetry at high temperatures each of the vectors $\vec \pi_i$ can freely rotate in the internal space. This symmetry, however, is valid only before the formation of the internal tetrahedron and is broken to $A_4 + S ( S_4 - A_4)$ when the tetrahedron is formed and the $\vec \pi_i$ are fixed to their position in fig. 1. In the language of many-particle physics fig. 1 is a frustrated antiferromagnet configuration\cite{frust}, because the spin vectors try to avoid each other as far as possible, but do not achieve to form a completely anti-parallel configuration.

Note that this breaking as yet has nothing to do with the spontaneous breaking of $SU(2)_L$, but is dictated by the internal dynamics which leads to the formation of one tetrahedral 'molecule'. Rather it can be related to the breaking of the so called 'custodial' SU(2) to be defined below.

As shown in section 3, the breaking of internal parity Q is accompanied by a breaking of parity in physical Minkowski space. The point is that assuming a universal field theory for 1+3+3 dimensions a connection will be established between the internal and external parity operations. Any particle with chiral interactions in the internal space will experience an internal polarization due to the chiral structure in fig. 1, and this polarization will be accompanied by a corresponding chiral interaction of the particle in the base space, an effect which will eventually be used to explain the $V-A$ structure of the weak interactions.

Within the formalism of section 3 the simultaneous violation of internal and external parity will show up in the simultaneous appearance of $\vec \tau$ and $\gamma_5$ in eqs. (\ref{multi1}) and (\ref{ar47}), where $\vec \tau$ denotes the triplet of internal Pauli matrices and $\gamma_5=i\gamma_1\gamma_2\gamma_3\gamma_4$ the $\gamma_5$-matrix in Minkowski space. These quantities are representatives of parity violating behavior in their respective spaces (internal $R^3$ and Minkowski space), because $\gamma_5$ gives it a pseudoscalar behavior in Minkowski space and $\vec \tau$ a pseudovector behavior in the internal space. They are the building blocks for the chiral spin vectors, which will now be constructed.  Namely, one chooses to define
\begin{equation} 
\vec\pi = \frac{1}{\Lambda^2} (\bar \psi i \gamma_5 \vec \tau \psi)
=\frac{2}{\Lambda^2} 
\begin{pmatrix}
- \operatorname{Im} [\bar D_R U_L- \bar D_L U_R]\\
\operatorname{Re} [\bar D_R U_L - \bar D_L U_R]\\
- \operatorname{Im} [\bar U_L U_R + \bar D_R D_L]
\end{pmatrix}
\label{multi1}
\end{equation}
where $\Lambda$ at this point is just a mass scale to keep the dimensions right. 
To make the list of components of the Higgs doublet eq. (\ref{hig77}) complete we write
\begin{eqnarray} 
\sigma &=& \frac{1}{\Lambda^2} (\bar \psi \psi) = \frac{1}{\Lambda^2} [\bar U U + \bar D D] 
         = \frac{2}{\Lambda^2} \operatorname{Re} [\bar U_L U_R + \bar D_R D_L] 
\label{multi2}
\end{eqnarray}
When combined to the Higgs potential eq.(\ref{hipo2}), the theory is invariant under $SU(2)_L\times SU(2)_R \times U(1)$ transformations, where the charge of the $U(1)$-transformations $\psi \rightarrow e^{i\alpha}\psi$ can be identified with the internal fermion number.
The vev of the $\sigma$ field 
\begin{eqnarray} 
\langle \sigma \rangle  = \frac{1}{\Lambda^2}  \langle \bar \psi \psi \rangle  
\label{multi133}
\end{eqnarray}
breaks this symmetry to $SU(2)_V \times U(1)$, where $SU(2)_V$ is the diagonal so called 'custodial' SU(2) group.
In the framework of the present model it can be identified with the internal spin SU(2) introduced before, and one concludes that although it is a symmetry of the Higgs potential it is not a symmetry of the system as a whole, because it is broken by the formation of the internal tetrahedron.


\begin{figure}
\begin{center}
\epsfig{file=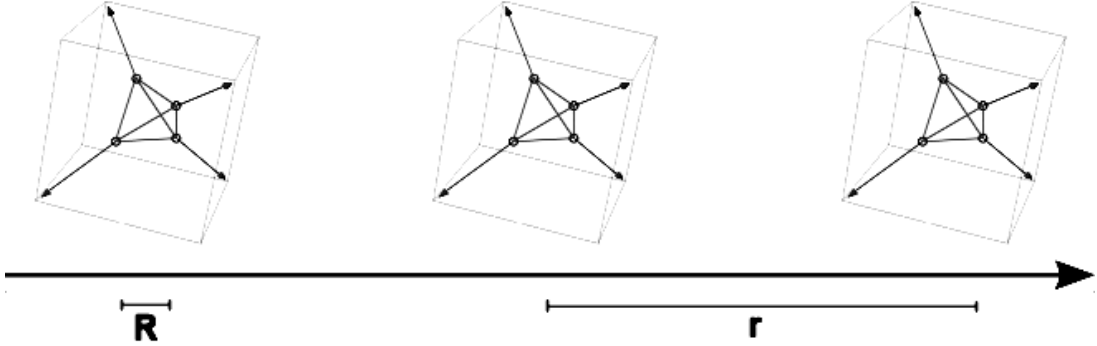,height=5.5cm}
\bigskip
\caption{The global ground state of the model after SSB consists of an aligned system of chiral tetrahedrons over Minkowski space (the latter is represented by the long arrow). R is the magnitude of a tetrahedron and r the distance between two of them. Associated to the 2 length scales R and r are the energy scales $\Lambda_R$ and $\Lambda_r$, as defined in the text. Before the SSB the chiral tetrahedrons are oriented randomly (not shown) and there is a corresponding local $SO(3)$ symmetry, because each rigid tetrahedron can be rotated freely and independently from the others. For reasons described in the main text, the covering group of this SO(3) can be identified with the group $SU(2)_L$, which gets broken spontaneously by the condensate $\langle \bar U U +\bar D D \rangle$. Though formally it can be considered as part of an SO(4) vector in the (2,2) representation of $SU(2)_L \times SU(2)_R$, physically it corresponds to a particle density, a scalar quantity giving the density of pairs of ionized tetrahedral components whose dynamical formation is responsible for the SSB at distances of order $\Lambda_F$, as explained in the main text.}
\nonumber
\end{center}
\end{figure}

Inserting (\ref{multi1}) and (\ref{multi2}) in (\ref{hipo2}), the bilinear term $\sim H^+ H$ of the Higgs potential has precisely the form of a 4-fermion interaction as appears in the Nambu-Jona-Lasinio (NJL) Lagrangian\cite{njl}
\begin{eqnarray}
L_{NJL}=\bar \psi (i\gamma_\mu  \partial^\mu -m) \psi+\frac{1}{\Lambda_R^2} [ (\bar \psi \psi)^2+(\bar \psi i \gamma_5 \vec \tau \psi)^2] 
\label{njl66}
\end{eqnarray}
where $m$ denotes the bare mass of the fundamental fermions $\psi=(U,D)$ and $\Lambda_R^{-2}$ the NJL-coupling which for dimensional reasons is written in terms of a new scale $\Lambda_R$. In technicolor theories this scale is usually interpreted as the mass of a heavy vector boson exchanged between the techniquarks, and is running due to renormalization group effects. Introducing a vev $\langle \bar \psi \psi \rangle$ a comparison between (\ref{njl66}) and (\ref{hipo}), i.e. 
\begin{eqnarray}
\mu^2 H^+ H = \frac{1}{\Lambda_R^2} [ (\bar \psi \psi)^2+(\bar \psi i \gamma_5 \vec \tau \psi)^2] 
\label{njxxx}
\end{eqnarray}
fixes the unknown energy scale $\Lambda$ in eqs. (\ref{multi1})-(\ref{multi133}) in terms of $\Lambda_F$ and $\Lambda_R$. Renormalization effects within the NJL-model even allow to derive a gap equation for the mass of the fundamental fermion. For consistency reasons, at low energies $\sim \Lambda_F$ all scales involved $\Lambda \sim \mu \sim \Lambda_R$ must then be of the same order O($\Lambda_F)$.

In contrast, at high energies, where there is no condensate and no symmetry breaking potential ($V > 0 \rightarrow L<0$), the NJL coupling $\Lambda_R^{-2}$ must be small and negative, the scale $\Lambda_R$ in the present model
roughly corresponding to the extension of an internal tetrahedron, cf. fig. 2. In that regime it is thus a repulsive potential and leads to the antiferromagnetic configuration fig. 1. If one is looking closely, one can identify the $\vec \pi \vec \pi$ term in the original Higgs potential eq. (\ref{hipo2}) together with (\ref{multi1}) as a sort of an internal Heisenberg spin-spin interaction. The Hamiltonian of such an interaction takes the form
\begin{eqnarray} 
H_H =-J \sum_{i\neq j} \vec \pi_i \vec \pi_j 
\label{hei9}
\end{eqnarray} 
where the sum is over sites i and j of a given discrete structure and J is the coupling derived from an exchange integral in internal space. $J>0$ accounts for ferromagnetic attraction and $J<0$ for antiferromagnetic repulsion. The appearance of an exchange integral is a quantum effect due the Pauli principle and explains the phenomenon of magnetism in solid state physics. In the present model J is the internal exchange integral defined by integrating over internal coordinates. A precise definition of J and a complete description of the connection between the NJL-model and an $SU(2)_L \times SU(2)_R$ Heisenberg type of spin interaction will be given at the end of section 3.

\begin{figure}
\begin{center}
\epsfig{file=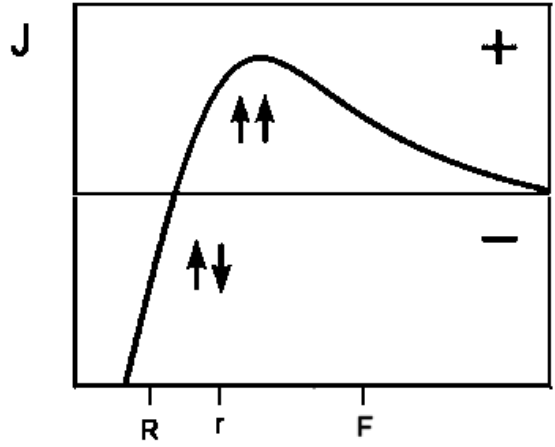,height=7cm}
\caption{Bethe-Slater curve: the exchange integral J as a function of the distance between 2 internal spin vectors. If the spin vectors lie within one tetrahedron, their distance is small $\sim R$ and according to the figure J is negative. This corresponds to antiferromagnetic behavior and leads to the formation of the frustrated structure fig. 1 with symmetry $A_4 + S ( S_4 - A_4)$, because the spin vectors try to avoid each other as far as possible. In contrast, if the internal spin vectors belong to different tetrahedrons, their distance is large, of order r, and J is positive. This corresponds to ferromagnetic behavior. At distances of the order of the Fermi scale, in the picture denoted by F, one is still in the ferromagnetic regime. The reader is warned however that in that region a further long range correlation comes into play which is responsible for the SSB, as described in the main text. It may be remarked, that in ordinary magnetism the Bethe-Slater curve is used to understand the magnetic behavior of metals. Elements like Fe and Co are characterized by large lattice spacings and corresponding large distances between spin vectors, much larger than the extension of the electron wave function. In these cases one has $J>0$ and a ferromagnetic behavior. On the other hand, antiferromagnets like Cr and Mn are characterized by small lattice spacings and corresponding small distances between spin vectors, typically not much larger than the extension of the electron wave function. In these cases $J<0$, i.e. antiferromagnetic behavior.}.
\nonumber
\end{center}
\end{figure}

Comparing (\ref{hei9}) with (\ref{hipo2}) one can identify $J=\mu^2/2$, i.e. there is attraction between the internal spin vectors in the SSB regime of energies $\sim \Lambda_F$, where $\mu^2 > 0$. In contrast, for high energies (small distances) where the SSB gets lost, the potential is repulsive with $J<0$ and leads to the internally frustrated antiferromagnetic configuration of fig. 1. Such an energy dependence of the exchange integral is very well known from the theory of magnetism and is given by the so-called Bethe-Slater curve depicted in fig. 3. It should be noted however, that in the context of normal magnets, the Bethe-Slater curve is derived from 3-dimensional exchange integrals, while here at energies $\sim \Lambda_F$ we are facing a 6-dimensional problem. Anyway, in the more sophisticated model described below further interactions come into play which account for the SSB and also contribute to the attraction at large distances.

In the low energy (SSB) regime the $\vec\pi-\vec\pi$ interaction eq. (\ref{hei9}) seems to disappear, because the sum of terms $\sim \vec\pi^2$ vanishes in the potential 
\begin{eqnarray}  
V(H)= -\frac{1}{2}\mu^2 [(\Lambda_F+\phi)^2 +\vec\pi^2]+\frac{1}{4}\lambda [(\Lambda_F+\phi)^2 +\vec\pi^2]^2
\label{hipo2555}
\end{eqnarray}
However, when the $\vec \pi$ triplet is absorbed as the longitudinal mode of the $\vec W$-boson the internal Heisenberg spin interaction reappears as part of the mass term $m_W^2 W_\mu W^\mu$.

The internal antiferromagnetic repulsion at short distances follows from the internal Coulomb force which governs the exchange integral J. A rigorous proof of this statement will be given in section 3, where a field theory will be defined which generates the internal Coulomb force.
At this point, we can calculate the energy for the local vacuum state fig. 1 and prove that it is a local minimum. To see this, just consider the products $\vec \pi_i \vec \pi_j= |\vec \pi|^2 \cos \alpha_{ij}$ where $|\vec \pi|^2$ is the length of the spin vectors and $\alpha_{ij}$ the angle between them ($i,j=1,2,3,4$). It can then easily be seen that for the configuration fig. 1 one gets the same energy as for the ideal antiferromagnetic configuration where 2 spin vectors show in the +z and the other 2 in the -z direction, namely $\sum_{i\neq j=1}^4\vec \pi_i \vec \pi_j=2 |\vec \pi|^2$, while all other configurations give larger values.

When the distances become larger and the energy is lowered towards the Fermi scale, J changes sign due to the Bethe-Slater effect shown in fig. 3, and would in principle lead to an attractive 'ferromagnetic' interaction between 2 distinct tetrahedrons fig. 2, so that an alignment of spin vectors of these tetrahedrons would occur, induced by the last term $\sim \vec \pi \vec \pi$ in the NJL-Lagrangean eq. (\ref{njl66}). A useful order parameter for magnetic systems in such a situation is the total magnetization, in our case the sum of all internal chiral spin vectors over internal and Minkowski space. Unfortunately, in the present case the total internal spin vector is not suitable to use. The point is that for a single local ground state configuration fig. 1 the 'magnetization' vanishes:
\begin{eqnarray} 
\vec \pi = \sum_{i=1}^4 \vec \pi_i = 0
\label{tomag}
\end{eqnarray}
This is simply due to the tetrahedral arrangement of the chiral spin vectors and implies that effectively there is no internal magnetic interaction between 2 tetrahedrons. In other words, looked at from the distance the 'magnetic field' of a single internal tetrahedron cannot be perceived. This is the deeper reason why there are no chiral $\pi$-condensates in the Standard Model, cf. eq. (\ref{hvev}). One has to search for another order parameter, and that is how the Higgs doublet H comes into play. According to eq.(\ref{hig77}), H contains besides the chiral spinvector $\vec \pi$ the scalar field $\sigma$, and it is this quantity which carries the condensate and should be used as the order parameter, cf. eq.(\ref{hvev}). The deeper reason for that is that at energies $\sim \Lambda_F$ particle-antiparticle interactions come into play which induce the long range correlation underlying the dynamics of the SSB. Details will be explained below in this section.

Strictly speaking one must distinguish the chiral spin vector $\vec \pi$ for the local internal ground state in fig. 1, which sums up to zero, from the fields $\vec \pi$ in the Higgs doublet, which can be interpreted as the longitudinal modes of the W/Z bosons. Conceptually, they are related to each other in the same way as the vacuum condensate $\langle \sigma \rangle$ is related to the Higgs field $\phi$. While the spin vectors can be defined for one tetrahedron alone (just as in ferromagnetism the spin vector $f^+ \vec \tau f$ can be defined for one electron alone), the bound states, when formed, turn out to be extended objects over many tetrahedrons over Minkowski space.

To summarize the situation, the breaking of the internal symmetries consists in 2 steps: 
\begin{itemize}
\item The formation of a tetrahedron due to an internal interaction within one single internal space. This interaction is 'antiferromagnetic' and leads to a 'frustrated' configuration, because the spin vectors try to avoid each other but do not achieve to form a completely anti-parallel configuration. How this kind of internal magnetism can be understood from a more fundamental higher dimensional theory will be explained in section 3. The frustrated tetrahedron breaks internal spin SU(2) as well as internal parity to the Shubnikov group $A_4+S (S_4 - A_4)$. This symmetry breaking however is {\it not} spontaneous but arises from the arrangement of a single 'molecule' due to the internal antiferromagnetic exchange interaction which avoids parallel spin states. The local ground state thus is a chiral configuration, i.e. it violates internal and, as shown in section 3, external parity, and the whole system is left $SU(2)_L$-symmetric - where the precise definition of the group $SU(2)_L$ is as follows:
\item Before the SSB each local tetrahedral ground state can rotate independently of the others, i.e. it can freely rotate as a rigid body over its base point in Minkowski space, and this rotational symmetry of the rigid chiral spin vector system corresponds to a $SO(3)$ symmetry group, whose covering group is taken to define $SU(2)_L$.
As a matter of fact it is a local symmetry, because the rotation can be different for tetrahedrons over different base points. Furthermore, due to the $V-A$ structure of the interactions induced by the tetrahedral structure fig. 1, it is a symmetry involving only left handed particles. This issue will be worked out in detail in the next section.
\end{itemize}
Corresponding to this scenario two new energy scales may be introduced: one is the magnitude of the tetrahedron $\Lambda_R$, which determines the strength of the NJL-coupling at small distances, and the other the average distance $\Lambda_r$ between 2 tetrahedrons in Minkowski space (cf. fig. 2). 
One can also associate these scales to 2 temperatures $T_R > T_r$. Cooling down the universe from big bang temperatures, at about $T_R$ the rigid tetrahedrons are formed in the internal fibers. Afterwards, at $T>T_r$, the spontaneous symmetry breaking sets in. In this regime of distances the NJL-coupling becomes positive, increases 
and finally, at the Fermi temperature, reaches its maximum value. In this picture the Fermi scale extends over many tetrahedrons, and it is well possible that an additional long range correlation is at work here, similar to the role phonons play in superconductivity. 

A tentative proposal will now be presented that provides such a long range correlation. Namely, consider an internal tetrahedron of spin vectors fig. 1, which from its very nature is an internal-particle antiparticle conglomerate and assume that for reasons of temperature it can 'ionize', i.e. separate into its particle and antiparticle component, and furthermore that the components can leave their original place in the crystal of tetrahedrons fig. 2 and can freely move around. There may in fact be a permanent equilibrium of ionizing and recombining tetrahedrons in the system, taking care that the structure of the global vacuum percolates over the whole of Minkowski space. In such a scenario the appearance of unwanted domain walls in the cosmological system is naturally suppressed.

It is further assumed that all on-site, fixed tetrahedrons are effectively neutral. In contrast, the ionized components, although in part screened by the bulk of on-site tetrahedrons, will attract and encircle each other even at distances as large as $\Lambda_F$ and thus temporarily form a sort of 'positronium' state, where the Fermi scale plays the role of its 'Bohr radius' and sets the length scale of the problem, the physics being controlled by momentum exchange processes of order $\Lambda_F$. In this weak binding limit, i.e. when the screening from the bulk is strong and the Bohr radius is much larger than the lattice spacing r, the interaction will be dominated by the long range part of the 'Coulomb' potential induced by the internal photon, as introduced and discussed in section 3. The attractive force is then described by the attractive version of the NJL-Lagrangian. If the number of pairs exceeds a certain critical density, a Bose-Einstein condensation of $\langle \bar \psi \psi\rangle$ sets in, producing the required phase transition. Through the SSB, the Higgs field comes into being, a coherent state corresponding to the superposition of all ionized and weakly bound pairs.


The expert reader will find that the start-out formulas eqs. (\ref{multi1})-(\ref{njxxx}) of these considerations are similar to what one has in simple technicolor models\cite{technireview}. It should be noted, however, that there are some important differences: first, there is no need of a (techni)color quantum number here, because the present model describes a strongly correlated fermion system in the sense of solid state physics and the bound states are formed by these correlations instead of by (techni)color forces. Furthermore, the fermions U and D do not interact directly with quarks and leptons\cite{lamm1}, and so the model does not have problems with FCNCs. Thirdly, in technicolor theories the value of the condensate is usually assumed to be 
\begin{eqnarray}
\langle \bar \psi \psi \rangle  \sim \Lambda_F^3
\label{hve33}
\end{eqnarray}
In other words, the extension of the condensates (and of the Higgs particle) is of the order of the Fermi scale. Such a value of the condensate is also appropriate in the present model, although the interpretation is somewhat different (see above).

As for the symmetry breaking, the usual technicolor argument goes as follows: before the introduction of gauge interactions the model has a global $SU(2)_L \times SU(2)_R$ invariance, and this symmetry is broken to custodial $SU(2)_V$ by the Higgs potential. However, when the SM gauge interactions are turned on, the $SU(2)_R$ part of the chiral symmetry group automatically disappears, so that the SSB concerns only the remaining $SU(2)_L$ factor. In the present model the disappearance of $SU(2)_R$ has a clear physical interpretation. It is simply due to the fact that the chiral tetrahedrons are formed. As proved in the next section, this formation implies the $V-A$ structure of the weak interactions.

It is interesting to note that at energies $ \leq \Lambda_F$ the microscopic tetrahedral structure does not shine up at all in the Higgs system. The only effect of the tetrahedral structure at low energies are the multiplets of the vibrational and spin excitations eq. (\ref{eq833hg}), to be interpreted as the observed quark and lepton spectrum.

I repeat that the internal spin transformations are local in the sense, that the tetrahedrons can be rotated {\it independently} over different points of the base space (Minkowski space), and can therefore be used as a basis to define local gauge interactions. One has here a fiber space, each fiber with a discrete crystalline structure. Connections can be defined over the fibers, which give rise to the gauge fields. While the photon is a story of its own to be discussed in the next section, the explicit construction of the $SU(2)_L$ gauge fields may easily be sketched, because it is quite similar to the construction of the Higgs doublet. In fact they are also bound states of the fundamental fermions U and D and differ from the chiral internal spin vector field $\vec \pi$ only by their Lorentz behavior: 
\begin{eqnarray} 
\vec W = \frac{1}{\Lambda^2} (\bar \psi \gamma_\mu (1-\gamma_5) \vec \tau \psi)
\end{eqnarray} 
Due to the appearance of the factor $\vec \tau$ they are polarized in internal space by the internal chiral vacuum fig. 1, and, as shown in the next section, this polarization will carry down to give a handedness $\sim 1-\gamma_5$ in Minkowski space, providing the chiral $V-A$ nature of the weak interactions.   

\section{The background Scenery: QED in 7 and 8 Dimensions}


The interested reader may worry, what kind of dynamical framework can account for the phenomena described in the last section. The correlations between the internal ($\vec \tau$) and external ($\gamma_5$) axial structures are so intriguing that one is tempted to look for a unifying higher dimensional model which comprises all the necessary features. In view of the internal 'magnetic' effects described in the last section, an immediate suggestion is to consider QED-like interactions in a 6+1 dimensional space $R^{6+1}$ with SO(6,1) symmetry which fibers out to give the $R^3$ fibers depicted in figures 1 and 2. This fibration 
may be associated with the formation of the tetrahedrons (fig. 1)
at scale $\Lambda_R$, which span a 3-dimensional subspace of $R^{6+1}$ and may in fact be used to pinpoint what a fiber is. 
Namely, the fibers can be {\it defined} to be spanned by the tetrahedrons, while everything orthogonal is called Minkowski space. One then has
\begin{eqnarray}
SO(6,1) \rightarrow SO(3,1) \times SO(3)
\label{fu80}
\end{eqnarray}
While the base space is to be identified with physical Minkowski space and its SO(3,1) Lorentz group, a kind of non-relativistic physics with rotational SO(3) will take place in the fibers. 

As to the dynamics I suggest to consider 6+1-dimensional QED ('QED7') broken down to the above fiber space, and want to show how this splits into ordinary QED4 plus a non-relativistic electrodynamics in the fibers, i.e. an interaction which to some extent can be discussed within the so-called NRQED framework\cite{piso,labe}. A major difference as compared to ordinary NRQED will be the appearance of a chiral factor $\gamma_5$ which prevents the whole problem from being fully factorizable and furthermore relates internal and external chiralities. 

The main ingredients of QED7 are a SO(6,1) spinor field (the QED7-'electron') and a massless vector field (the QED7-'photon'). As for the fermion there is a single 8-dimensional spinor representation in SO(6,1) decomposing as\cite{slansky}
\begin{eqnarray}
8 \rightarrow (1,2,2)+(2,1,2)
\label{eq8}
\end{eqnarray}
under the fibration $SO(6,1) \rightarrow SO(3,1) \times SO(3)$. Here representations of $SO(3,1) \times SO(3)$ are denoted by a set of 3 numbers $(a,b,c)$, where $(a,b)$ are representations of the Lorentz group and $c$ is the dimension of a $SO(3)$-representation. For example, c=2 corresponds to a non-relativistic Pauli spinor in internal space, whose 2 spin orientations are identified with the SU(2) flavors U and D introduced in the last section. It should be noted that (1,2,2) and (2,1,2) are complex conjugate with respect to each other, so one is the antiparticle representation of the other. 

The QED7-photon transforms according to the fundamental 7-dimensional representation of SO(6,1) and decomposes as\cite{slansky} 
\begin{eqnarray}
7 \rightarrow (2,2,1)+(1,1,3)
\label{eq7}
\end{eqnarray}
i.e. into an ordinary QED4 photon which is a singlet under internal spin SU(2) plus an internal 3-dimensional vector potential to describe the internal interactions. 

The Lagrangian of QED7 resembles that of QED4
\begin{eqnarray}
L_{QED7}=-\frac{1}{4} F_{\mu\nu}F^{\mu\nu} + \bar \psi (i\Gamma_\mu D^\mu - m)\psi
\label{qed77}
\end{eqnarray}
where $\mu$ and $\nu$ run from 0 to 6, $\psi$ is the 8-dimensional spinor of eq. (\ref{eq8}), $\Gamma_\mu$ are the Dirac matrices in 6+1 dimensions and $D^\mu=\partial^\mu+ieA^\mu$ is the covariant derivative containing the 7-vector multiplet $A^\mu$ of eq. (\ref{eq7}). 

To make the decomposition $SO(6,1) \rightarrow SO(3,1) \times SO(3)$ explicit one should decompose the corresponding 6+1 dimensional Dirac algebra. The Dirac matrices of SO(6,1) are 8$\times$8 matrices and can be built up as tensor products of Pauli matrices\cite{rossbook} 
\begin{eqnarray}
\Gamma_0 = \tau_1 \otimes \tau_0 \otimes \tau_0  \\
\Gamma_1 = i\tau_2 \otimes \tau_0 \otimes \tau_0  \\
\Gamma_2 = i\tau_3 \otimes \tau_1 \otimes \tau_0  \\
\Gamma_3 = i\tau_3 \otimes \tau_2 \otimes \tau_0  \\
\Gamma_4 = \underline{ i\tau_3 \otimes \tau_3} \otimes \tau_1  \\
\Gamma_5 = \underline{i\tau_3 \otimes \tau_3} \otimes \tau_2  \\
\Gamma_6 = \underline{ i \tau_3 \otimes \tau_3} \otimes \tau_3  
\label{eqxx77}
\end{eqnarray}
where the first 2 columns on the rhs correspond to Lorentz SO(3,1) and the last column to internal SO(3). $\tau_0$ is the 2-dimensional unit matrix, so that the first 4 $\Gamma$-matrices $\Gamma_{0,1,2,3}=\gamma_{0,1,2,3}\otimes \tau_0$ give a set of Dirac matrices in Minkowski space. The last 3 $\Gamma$-matrices have the form $\Gamma_{4,5,6}=i\tau_3 \otimes \tau_3 \otimes \tau_{1,2,3}$, i.e. they are proportional to $\vec \tau$ in the internal space part. The interesting point to note here is the appearance of a common prefactor $i\tau_3 \otimes \tau_3$ in $\Gamma_{4,5,6}$, which is nothing else than the matrix $\gamma_5 = i\gamma_1 \gamma_2 \gamma_3 \gamma_4$ in Minkowski space. We thus have $\Gamma_{4,5,6}=\gamma_5 \otimes \tau_{1,2,3}$ and this will in fact lead to the anticipated appearance of products of the form $\gamma_5 \vec \tau$ in the internal interactions, which is responsible for the structure of the NJL-Lagrangian eq. (\ref{njl66}). As will shortly be seen, this makes sure that internal parity violating effects from the $A_4$ symmetry structure are passed down to Minkowski space. In more mathematical terms it can be related to the structure and existence of octonion multiplication, when 6+1 spacetime is assumed to be spanned by the octonion units $I$,$J$,$K$,$L$,$IL$,$JL$ and $KL$\cite{conway,kantor,lamm}.

Writing $A^\mu=(\tilde A^{\mu=0-3},\vec C)$ in eq. (\ref{qed77}) the separation of ordinary QED from the internal interaction can be made explicit
\begin{eqnarray}
\bar \psi \Gamma_\mu A^\mu \psi =  \bar \psi \gamma_\mu \tilde A^\mu \psi + \bar \psi  \gamma_5 \vec \tau \vec C \psi 
\label{ar46}  
\end{eqnarray}
Now that one has established products $\gamma_5 \vec \tau$ in the internal couplings, one can try to derive the parity violation of the weak interaction. In principle, the presence of such a coupling corresponds already to a parity violating behavior, both in internal and Minkowski space. However, for this to actually become perceivable, an additional appropriate 'chiral situation' has to be provided, again both in internal and Minkowski space. In Minkowski space this can be achieved, for example, by using polarized beams or if there is a second vertex with a $\gamma_5$-coupling in the Feynman diagram of the process. An analogous requirement must be met in the internal space. In other words, a configuration with a handedness must be present, in order to pick up a non-vanishing contribution from the axial coupling, and this in the case at hand is given by the local chiral ground state structure fig. 1. As a matter of fact, the non-relativistic circumstances of the internal $R^3$ space make it a similar situation as one has in optical activity of molecules, where in addition to a circularly polarized photon there must be a handed molecule in order to produce a non-vanishing effect.

In conclusion, the structure of the second term in eq. (\ref{ar46}) looks quite promising, because it corresponds to a chiral interaction in Minkowski space. Unfortunately, its magnitude is of order of the electromagnetic coupling and not large enough to explain the antiferromagnetism at small distances fig. 3. On the other hand we know since the time of Heisenberg\cite{heise}, that ordinary magnetism is purely an effect of the Coulomb interaction plus the Pauli principle, which lead to the exchange integral J. What is therefore missing in the above equation, is an internal scalar potential $C_0$ to provide for the Coulomb force.

To introduce such a field we restart by considering one more dimension, namely a space with SO(6,2) symmetry group instead of SO(6,1)and decompose it as
\begin{eqnarray}
SO(6,2) \rightarrow SO(3,1) \times SO(3,1) \rightarrow SO(3,1) \times SO(3)
\label{ar789}  
\end{eqnarray}
i.e. we allow for a separate dynamics and time evolution within the fibers. Afterwards however (second arrow in eq. (\ref{ar789})), the fibers are fixed to their base point in Minkowski space and become non-relativistic at rest (no boosts allowed) with symmetry group SO(3).   

Introducing this preface step makes some difference concerning the fermions, and affects the internal photon in the desired way. To see this, one should remember that SO(6,2) possesses three 8-dimensional spinor representations. Two of these are Weyl representations ($8_L$ and $8_R$) which build up a 16-dimensional Dirac field, just as in 3+1 dimensions a Dirac fermion can be built from two 2-dimensional Weyl representations. The third 8-dimensional spinor representation $8_V$ coincides with the fundamental 8-dimensional vector representation of SO(6,2). The fact that these SO(6,2) representations appear in 3 inequivalent forms is known as triality\cite{dixonbook}, a characteristic property of this group, which makes it very special indeed and again goes back to the existence of the division algebra of octonions. When one decomposes these representations according to (\ref{ar789}) one obtains\cite{slansky} 
\begin{eqnarray}
8_L &\rightarrow & (1,2,1,2)+(2,1,2,1) \rightarrow (1,2,2)+(2,1,2) \label{eq888L} \\ 
8_R &\rightarrow & (1,2,2,1)+(2,1,1,2) \rightarrow (1,2,2)+(2,1,2) \label{eq888R} \\
8_V &\rightarrow & (2,2,1,1)+(1,1,2,2) \rightarrow (2,2,1)+(1,1,3)+(1,1,1) \label{eq8880}
\end{eqnarray}
Here representations of $SO(3,1)\times SO(3,1)$ are denoted by (a,b,c,d), where the first 2 numbers (a,b) stand for representations of the Lorentz group, and (c,d) characterize representations of the internal SO(3,1). While (\ref{eq888L}) and (\ref{eq888R}) yield the same structure as eq. (\ref{eq8}) in the limit eq. (\ref{fu80}), the expression (\ref{eq8880}) does not agree with eq. (\ref{eq7}), but contains the desired singlet field (1,1,1). Note that (2,2,1,1) resp. (2,2,1) denotes the ordinary photon, while (1,1,2,2)$\rightarrow$ (1,1,3)+(1,1,1) is the (internally) non-relativistic decomposition of the internal 'photon'.

As for the fermion representation (1,2,2)+(2,1,2) in (\ref{eq888L}) and (\ref{eq888R}) there is one important physical difference as compared to the representation (\ref{eq8}) of SO(6,1). The point is that in SO(6,2) the nonrelativistic internal component of the antiparticle spinor (2,1,2) possesses an internal antiparticle property as compared to (1,2,2), which makes it behave like a nonrelativistic 'positron' with respect to the internal space as compared to a nonrelativistic 'electron' in the case of (1,2,2). This property is lacking in SO(6,1) (no internal antiparticles) and is of importance for the Bose-Einstein condensation proposed in section 2.

The SO(6,2) Dirac spinor is the sum of $8_L$ and $8_R$
\begin{eqnarray}
8_L+8_R \rightarrow (1,2,1,2)+(2,1,1,2)+(1,2,2,1)+(2,1,2,1)=(\underline{12+21},\underline{12+21}) \nonumber
\end{eqnarray}
where the underlined expression is a short form which makes it easy to understand, that it decomposes into a Dirac fermion in Minkowski space times a Dirac fermion in the internal space (as long as it is internally relativistic and not fixed to its base point). This object enters the QED8-Lagrangian which formally has the same structure as eq. (\ref{qed77}) with the indices now running from 0 to 7.

The Dirac matrices of SO(6,2) appearing in the QED8-Lagrangian will be called $G_\mu$ and are 16$\times$16 Matrices which can be written as tensor products of the form
\begin{eqnarray}
G_0 = \tau_1 \otimes \tau_0 \otimes \tau_0 \otimes \tau_0  \\
G_1 = i\tau_2 \otimes \tau_0 \otimes \tau_0 \otimes \tau_0  \\
G_2 = i\tau_3 \otimes \tau_1 \otimes \tau_0 \otimes \tau_0  \\
G_3 = i\tau_3 \otimes \tau_2 \otimes \tau_0 \otimes \tau_0  
\end{eqnarray}
\begin{eqnarray}
G_4 = \underline{ \tau_3 \otimes \tau_3} \otimes \tau_1 \otimes \tau_0  \\
G_5 = \underline{i\tau_3 \otimes \tau_3} \otimes \tau_2 \otimes \tau_0  \\
G_6 = \underline{ i \tau_3 \otimes \tau_3} \otimes \tau_3 \otimes \tau_1  \\ 
G_7 = \underline{ i \tau_3 \otimes \tau_3} \otimes \tau_3 \otimes \tau_2  
\end{eqnarray}
where the first 2 columns on the rhs correspond to the Lorentz group and the last 2 to the internal SO(3,1). Looking closely at these equations one understands that $G_{0,1,2,3}$ yield the ordinary Dirac matrices in Minkowski space (up to trivial factors $\tau_0 \otimes \tau_0$), and $G_{4,5,6,7}$ yield Dirac matrices in the internal space - however with a prefactor $\gamma_5 \sim \tau_3 \otimes \tau_3$.

Taking the non-relativistic limit in the internal fibers we end up with the $\gamma$-matrices discussed after eq. (\ref{eqxx77}). Furthermore, there is an internal scalar potential $C_0=A_4$ in addition to the internal vector potential $\vec C = A_{\mu=5,6,7}$. What we have achieved then is that we can apply the ordinary NRQED machinery to the internal spaces (except for the appearance of factors of $\gamma_5$ in the interactions). For example, the internal dynamics will be governed by the (slightly modified) Pauli Lagrangian
\begin{eqnarray}
L_{2f}=\bar \psi \{ i D_t + \frac{1}{2m}\vec D^2 + \frac{e}{2m}\gamma_5 \vec \tau \vec B \} \psi
\label{ar47}
\end{eqnarray}
where $m$ and $e$ are the mass and charge of the fundamental fermion $\psi =(U,D)$. $D_t=-\partial_t+ie\gamma_5 C_0$ 
and $\vec D =-\vec \nabla +ie\gamma_5 \vec C$ are covariant derivatives in the internal dimensions. $\vec B$ is the internal magnetic field strength of the internal photon, $C_0$ its desired scalar and $\vec C$ its vector potential.
The Pauli Lagrangian may be considered as the leading order NRQED contribution to the 2-fermion interactions of internal NRQED. There is also a leading 4-fermion Lagrangian\cite{piso,labe} which contains the terms arising in the NJL-Lagrangian (\ref{njl66}):
\begin{eqnarray}
L_{4f}= \frac{d_s}{m^2}(\bar \psi \psi)^2+\frac{d_v}{m^2}(\bar \psi  \gamma_5 \vec \tau \psi)^2
\end{eqnarray}
with $d_s, d_v =O(\alpha)$ being couplings of the effective NRQED field theory. They are obviously too small to account for the internal magnetic effects discussed in the last section. Strong correlations are needed to explain the formation of the internal tetrahedral structure. As discussed earlier, these naturally arise as exchange contributions via the (internal) Pauli principle.

To make this statement more precise we consider the situation before the formation of the tetrahedrons where the system still has its $SU(2)_L\times SU(2)_R$ symmetry. As well known, this group is isomorphic to an SO(4), whose fundamental representation (2,2) is spanned by the '4-vector' $T=(\sigma,\vec\pi)$, which is nothing else than the original Higgs doublet eq. (\ref{hig77}). The representation eq. (\ref{hipo2}) of the Higgs potential explicitly shows that it is invariant under this SO(4) group.

We now want to analyze the behavior of a pair of internal non-relativistic fermions, using the generalization of the Pauli principle to the internal space. The fermions called a and b are assumed to be bound on sites with internal coordinates $\vec q$ and $\vec r$. Such a pair will predominantly interact via the internal Coulomb potential $V=V(|\vec q -\vec r|)$ generated by the exchange of internal photons. We want to show that the net outcome of such an analysis is an SO(4) Heisenberg type of magnetic model which in the continuum limit agrees with the quadratic term in (\ref{hipo2}) or, equivalently, with the NJL-interaction term in eq. (\ref{njl66}).

Using the chiral properties of QED8, it can be shown that the internal spin part of the internal wave function for the a-b-system in the (2,2) representation is symmetric under the exchange $a\leftrightarrow b$. It then follows from the (internal) Pauli principle that the corresponding internal $R^3$ part of the wave function must be antisymmetric, of the form
\begin{eqnarray}  
f_{(2,2)}=\frac{1}{\sqrt{2}} [f_a(\vec q) f_b(\vec r) - f_a(\vec r) f_b(\vec q)]
\label{hei812}
\end{eqnarray}
corresponding to an energy eigenvalue $E_{(2,2)}=\langle f_{(2,2)}|V|f_{(2,2)}\rangle$.

Eq. (\ref{hei812}) may be compared to the internal $R^3$ wave function in case that an SO(4) singlet (1,1) is formed out of the pair a and b.
In this case a symmetric $R^3$ part arises, because the internal spin part is antisymmetric, and the energy can be calculated via $E_{(1,1)}=\langle f_{(1,1)}|V|f_{(1,1)}\rangle$. Working out the formulas one finds for the energy difference 
\begin{eqnarray}  
E_{(2,2)}-E_{(1,1)}=2J=2\langle a(\vec q) b(\vec r)|V|a(\vec r) b(\vec q)\rangle
\label{hei81a}
\end{eqnarray}
which defines the internal exchange integral J. If one now parametrizes this energy spectrum in the Heisenberg manner as product of 2 spin 4-vectors $T_a$ and $T_b$
\begin{eqnarray}  
H_H=A+BT_aT_b
\label{hei815}
\end{eqnarray}
one easily finds $B=-J$, while the constant A is irrelevant for most considerations.

Clearly, the products $T_aT_b$ are the discrete version of the term $\sim \sigma^2 + \vec \pi^2$ in the Higgs potential and in the NJL-Lagrangian eq. (\ref{njl66}). In other words, the strong (anti)binding effects which lead to the frustrated tetrahedral structure fig. 1 arise in a similar manner from the $SU(2)_L \times SU(2)_R$ symmetry structure as the Heisenberg-model in ordinary magnetism from the underlying spin SU(2). 

Numerically it turns out that $J<0$. This sign of the exchange integral corresponds indeed to the antiferromagnetic behavior needed at small distances $\sim \Lambda_R$ for the formation of the internal 'frustrated' tetrahedral structure fig. 1. As argued in section 2, after this formation there is a local $SU(2)_L$ symmetry left in the basket. To account for the attractive SSB forces, which lead to the breaking of this symmetry at large distances $\sim \Lambda_F$, in principle integrals over the whole of 6-dimensional space have to be evaluated, and furthermore the long range correlation picture with the 'ionization' of tetrahedrons and condensation of pairs as discussed in section 2 has to be taken into consideration.

\section{Conclusions}

In this paper the SM Higgs mechanism has been analyzed on the basis of dynamics taking place in a 3-dimensional internal space with a chiral tetrahedral structure. It was shown that weak parity violation can be completely understood from interactions within one single tetrahedron and has no spontaneous character. In contrast, the breaking of $SU(2)_L$ is spontaneous and due to an alignment of {\it all} internal fibers over the whole of Minkowski space. 

Since this is a rather unusual approach it may seem hard to understand where it comes from and to where it will lead, in particular because I have mostly restricted myself to the symmetry breaking aspects and did not consider other questions\cite{lamm,lamm1}. Actually, there is a certain physical picture in my mind where our universe resembles a huge crystal of molecules, each 'molecule' of tetrahedral form like in fig. 1, and arranged in such a way that certain symmetries are (spontaneously) broken. For such a model to be consistent, a 6+1 dimensional space time has been introduced, i.e. the 'molecules' extend to 3 internal dimensions which are orthogonal to physical space. The strong correlations within this system provide for the Higgs particle and the weak vector bosons as bound states. Furthermore, as shown in ref. \cite{lamm1}, internal spin and vibrational excitations can be interpreted as the quark and lepton spectrum. Then, it happens that an excitation in one molecule is able to excite an excitation in the neighbouring internal space and thus can travel as a quasi-particle through Minkowski space with a certain wave vector $\vec k$ which is to be interpreted as the physical momentum of the particle. 

These considerations directly lead to a comment on the widespread belief that internal spaces must be 'compactified' to a small radius because otherwise they would be easily detectable. Such an assumption is not really needed in the present model, because the tetrahedrons arrange themselves to form a crystal which exists only on a 3-dimensional subspace of the 6-dimensional universe, and everything physically relevant for us happens in a small neighbourhood of this 3-surface. The point is that an internal tetrahedron fig. 1 cannot be extended to a crystal structure in the internal fibers. The reason for that is that a crystal with {\it point} symmetry $A_4+S(S_4-A_4)$ cannot grow into the internal dimension, because there is no 3-dimensional {\it space} symmetry group possessing this point symmetry\cite{nospace}. Once a tetrahedron or its constituents ionize from the 3-surface crystal, they cannot dock at an arbitrary point above or below the surface, but have to search for a point defect within the 'surface' crystal, where they fit in. As long as they cannot find such a vacancy they fluctuate around near the surface forming the positronium like 'Cooper' pairs introduced in section 2.



In ferromagnets with Pauli spinors $f=(f_\uparrow,f_\downarrow)$ the appropriate order parameter is the sum of the spin vectors 
\begin{eqnarray}
<f^+ \vec \tau f>
\label{g3456}
\end{eqnarray}
whereas in superconductivity the condensate of Cooper pairs 
\begin{eqnarray}
<f_\uparrow f_\downarrow> 
\label{h3456}
\end{eqnarray}
determines the order of the system. The phenomena of high energy physics are such that the relevant quantity is the (internal + relativistic) generalization of (\ref{h3456}), but not that of (\ref{g3456}). In section 2 an intriguing explanation was found for this fact. Nevertheless, it was shown that the chiral spin vectors eq. (\ref{multi1}) play an important role in the dynamics of the system. They are not only essential ingredients of the Higgs doublet and the NJL Lagrangian, but in the internal spaces they interact in an 'antiferromagnetic' way, thereby determining the local ground state of the system.

I have finished this paper leaving a lot of open questions. For example, the mixing of the photon and Z-boson has not been worked out. Then there is the question, whether the fundamental fermions U and D are in principle observable or whether they act just as a sort of background fields for the physical excitations. Thirdly, it is unclear, whether the condensate $\langle \bar \psi \psi \rangle = \langle \bar U U + \bar D D \rangle $ is really $SU(2)_V$-symmetric or whether $\langle \bar U U \rangle \neq \langle \bar D D \rangle $. This is a well justified question in view of the fact that the chiral tetrahedron breaks $SU(2)_V$. And finally, there is the question, which force keeps the original tetrahedron (the circles in fig. 1) together.

\section{Appendix}

As kindly suggested by the referee I would like to add some comments about the possible interplay of my model with cosmology and gravity. First, I will try to re-develop the history of the early universe within the assumptions of the present framework. Based on these arguments I will afterwards discuss the nature of the gravitational force.

At big bang temperatures there were the free fermions $\psi$ floating around as a Fermi gas in $R^{6+1}$ space at extremely high pressure and temperature. While the universe was cooling down 3 important phase transitions happened: 

I. the formation of tetrahedral molecules from the fundamental fermions $\psi$ at very high temperature $T\approx \Lambda_R $, where the scale $R$ is roughly given by the extension of one molecule. Although this process is not a phase transition in the strict sense it has certainly released a large amount of energy which has amplified the initial expansion of the universe. Note that with 4 molecular sites each molecule 'fills' only 3 of the 6 spatial dimensions.

II. the formation of a 3+1-dimensional 'hyper-crystal' within $R^{6+1}$ built from tetrahedral molecules took place at somewhat lower temperatures $T \approx \Lambda_r$, where $r$ is roughly given by the distance between 2 molecules. This alignment of all tetrahedral structures is a coordinate alignment and to be distinguished from the isospin vector alignment (item III) describing the electroweak phase transition. It puts all 3-dimensional molecular structures in parallel thus separating an internal 3-dimensional space from the rest. In other words, the crystal expands into a 3+1-dimensional subspace of $R^{(6,1)}$, while the tetrahedrons extent into what becomes the 3 internal dimensions. The natural scale for this phase transition is the distance r between tetrahedrons as introduced in fig. 2. As a crystallization process it is a first order phase transition associated with the sudden release of a large amount of energy, and one may speculate whether it is related to cosmic inflation.

III. the arrangement of isospins at the Fermi scale $T\approx\Lambda_F$. At higher temperatures the isospin vectors have fluctuated randomly with an associated internal 'Heisenberg' $SU_L(2)$ symmetry but at $\Lambda_F$ they arrange into the chiral isomagnetic structure figs. 1 and 2. At that point the so far freely rotatable internal spins got ordered and $SU_L(2)$ was broken to the Shubnikov group. Note that $SU_L(2)$ is a local symmetry because isospins can be rotated separately over each point of the Minkowsky base space.

Since II happens after I, i.e. at lower temperature, one naturally expects $\Lambda_R > \Lambda_r$ (i.e. $R < r$) in agreement with fig. 2 and the Bethe-Slater curve fig. 3. Both scales are much larger than the Fermi temperature $\Lambda_F$ at which the isospins align. The point is that while $\Lambda_F$ is defined as the critical point of transition III, the values of the exchange integrals and therefore the isomagnetic behavior are determined by r and R.

From the cosmological point of view the transition II is the most interesting because it describes the formation of the crystal, i.e. of our world. Therefore, one may try to understand the connection of the present discussion with the phenomenon of gravity. Unfortunately, it is not clear what the relation between the scales $\Lambda_R$, $\Lambda_r$ and the (Planck) scale of gravity is. Furthermore, it must be noticed that a rigid spatial lattice over Minkowski space like insinuated in fig. 2 is inconsistent with the principle of equivalence. 

At this point I can only make some rather rudimentary remarks about the way to solve this problem. As well known, in general relativity mass and energy lead to a curvature of the spacetime continuum. Sakharov was one of the first to interpret this on the basis of metric elasticity\cite{sak,add1,gron,bomm}. Furthermore, he thought gravity forces might be explainable from a microscopic structure which in some sense is analogous to the molecular structures responsible for material elasticity in low-energy physics. In this spirit I want to interpret the hyper-crystal as a {\it discrete micro-elastic spacetime continuum} - while maintaining the rigid tetrahedral structure within the internal spaces. In other words, while the tetrahedral molecules are rigid bodies which align in their internal spaces, the nature of the inter-tetrahedral {\it coordinate} interactions is elastic. These elastic forces between tetrahedrons are remnants of the original interactions of the fundamental fermions, a feature that has to be added to the list of necessary properties that the fundamental fermion interaction has to fulfill.



\begin{thebibliography}{99}
\bibitem{wein1} S. Weinberg, Phys. Rev. D19 (1979) 1277.
\bibitem{suss1} L. Susskind, Phys. Rev. D20 (1979) 2619.
\bibitem{lamm} B. Lampe, arXiv:1201.2281 [hep-ph] (2012), Int. J. Theor. Phys. 51 (2012) 3073.
\bibitem{lamm1} B. Lampe, arXiv: arXiv:1212.0753 [hep-ph] (2012), Mod. Phys. Lett. A28 (2013) 135.
\bibitem{dixonbook} G.M. Dixon, Division Algebras, Kluwer Books, 2009. 
\bibitem{conway} J. Conway and D. Smith, On Octonions and Quaternions, 
Peters Publishing, Natick, MA (2003). 
\bibitem{kantor} I. L. Kantor and A. S. Solodovnikov, Hypercomplex Numbers -
an Elementary Introduction to Algebras, Springer Verlag, Berlin, 1989.
\bibitem{alta} for a review of various other $A_4$-type theories see G. Altarelli, arXiv:hep-ph/0508053 [hep-ph] (2005).
\bibitem{technireview} for a review on technicolor-like theories see R. Contino, The Higgs as a Composite Nambu-Goldstone Boson, in TASI Lectures 2009, eds. C. Csaki and S. Dodelson, World Scientific, 2011.
\bibitem{njl} J.O. Andersen, L.T. Kyllingstad, J. Phys. G37 (2010) 19.
\bibitem{slansky} R. Slansky, Phys. Rep. 79 (1981) 1.
\bibitem{frust} J. Schnack, Dalton Trans. 39 (2010) 4677, arXiv:0912.0411v1 [cond-mat] (2009). 
\bibitem{piso} A. Pineda and J. Soto, Phys. Rev. D59 (1998) 16.
\bibitem{labe} P. Labelle, Phys. Rev. D58 (1998) 93.
\bibitem{heise} W. Heisenberg, Z. Phys. 49 (1928) 619. 
\bibitem{shub} A. P. Cracknell, Progr. Theor. Phys. 35 (1966) 196.
\bibitem{borov} A. S. Borovik-Romanov and H. Grimmer, International Tables for Crystallography D (2006) 105.
\bibitem{bata} M. El-Batanouny and F. Wooten, Symmetry: a computational approach, Cambridge University Press, 2008.
\bibitem{rossbook} G. G. Ross, Grand Unified Theories, Oxford University Press, 1984.
\bibitem{nospace} Kim, S. K., Group theoretical methods and applications to molecules and crystals, 1999, Cambridge University Press, ISBN 978-0-521-64062-6.
\bibitem{sak} A. D. Sakharov, Doklady Akad. Nauk S. S. R. 177, 70-71 (1987), Sov. Phys. - Doklady 12, 1040-1041 (1968).
\bibitem{add1} S. L. Adler, Rev. Mod. Phys. 54, 729 (1982).
\bibitem{gron} F. Gronwald and F.W. Hehl, in Advances in Modern Continuum Dynamics, Isola d’Elba, 1991.
\bibitem{bomm} C.G. Bohmer and N. Tamanini, arXiv:1301.5471v2 [gr-qc] (2013)
\end{thebibliography}
\end{document}